\documentclass[10pt,letterpaper]{article}
\usepackage{opex3,cite}

\begin{document}

\title{Polarization spectroscopy and magnetically-induced dichroism of the potassium D$_2$ lines}

\author{K. Pahwa, L. Mudarikwa, and J. Goldwin$^\ast$}

\address{Midlands Ultracold Atom Research Centre, School of Physics and Astronomy, \\
			University of Birmingham, Edgbaston, Birmingham B15 2TT, UK}

\email{$^\ast$j.m.goldwin@bham.ac.uk} 



\begin{abstract}
We study modulation-free methods for producing sub-Doppler, dispersive line shapes for laser stabilization near the potassium  D$_2$ transitions at 767~nm. Polarization spectroscopy is performed and a comparison is made between the use of a mirror or beam splitter for aligning the counter-propagating pump and probe beams. Conventional magnetically-induced dichroism is found to suffer from a small dispersion and large background offset. We therefore introduce a modified scheme, using two spatially separated pump-probe beam pairs. Finally we compare our results to methods using phase modulation and heterodyne detection.
\end{abstract}



\section{Introduction}

The stabilization of laser frequencies near atomic resonances is a basic requirement for laser cooling and trapping experiments.  The required frequency stability is set by the natural atomic linewidth, which is on the order of a few MHz for the alkali atoms in common use.  A convenient way to obtain narrow frequency discriminants near resonance is through the use of sub-Doppler saturated absorption spectroscopy \cite{Han71}. This method is simple and robust, and can be performed with a room-temperature gas reference cell.  The transmission through the cell of a weak probe beam is recorded on a photodiode.  On its own, sweeping the frequency of this probe would result in a roughly Gaussian absorption feature, Doppler-broadened to a width of a few hundred MHz.  If a bright counter-propagating pump beam of the same frequency is introduced, the two beams are only simultaneously resonant for atoms with zero velocity along the optical axis.  For these atoms the pump beam reduces the ground-state population, and therefore the absorption, leading to a narrow Lorentzian feature known as a Lamb dip \cite{Ber11}.

Saturated absorption in this form is not directly suitable for laser frequency stabilization.  Typical servo control requires a so-called error signal which is proportional to changes in frequency in the vicinity of a zero-crossing which sets the lock point.  Although one can lock to the side of a saturated absorption feature by subtracting an electronic offset, the resulting signal is sensitive to variations in laser power and vapor pressure. A variety of methods have been developed for producing more robust error signals from atomic spectroscopy.  These can be classified broadly according to whether or not they require modulation of the laser phase, frequency or amplitude (alternatively a modulation may be applied to the atomic resonances \textit{via} the Zeeman effect in an oscillating magnetic field \cite{Din92}). Although there are a number of modulation-based methods, they all utilize phase-sensitive detection to generate dispersive error signals. On the other hand, modulation-free techniques have been developed, eliminating the need for radio-frequency electronics and modulators. Popular modulation-free methods include polarization spectroscopy \cite{Wie76,Pea02}; magnetically-induced dichroism, both Doppler-broadened \cite{Cor98} and sub-Doppler \cite{Shi99,Was02,Pet03}; and Sagnac interferometry \cite{Rob02,Wei10}.

Our interest is in stabilizing lasers near the D$_2$ transitions of potassium, at 767~nm.  In a previous publication we investigated the dependence of saturated absorption on vapor pressure and pump and probe intensities, and compared two types of modulation spectroscopy \cite{Mud12}.  Here we study polarization spectroscopy, and present a modified setup for magnetic dichroism.  These two methods were chosen due to their similar experimental arrangements, and their widespread use with other atoms, especially rubidium and cesium.  To the best of our knowledge, there has been no study of polarization spectroscopy with potassium, and only a single recent demonstration of magnetic dichroism which we became aware of while preparing this manuscript \cite{Pic12}. 

\begin{figure}[htbp]
\centering\includegraphics[width=10cm]{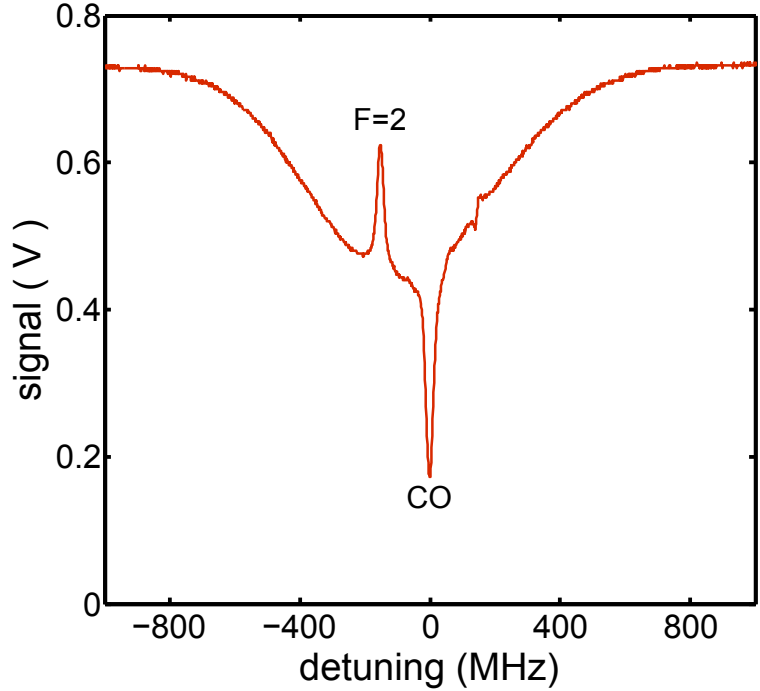}
\caption{\label{fig:01}Saturated absorption of potassium, taken at a cell temperature of $48~^\circ$C.  The labeled transitions are described in the text.}
\end{figure}

As described in detail in \cite{Mud12}, the hyperfine level splittings in potassium are much smaller than in rubidium and cesium, leading to a number of important spectroscopic differences. Since the natural half-width at half-maximum (HWHM) for potassium $\gamma=2\pi\times 3.017$~MHz \cite{Wan97} is on the order of the excited-state splittings, neighboring transitions tend to overlap and significant off-resonant excitation can occur.  Additionally, the isotope shifts and ground-state splittings are smaller than the Doppler width.  This leads to a single Doppler background for both naturally abundant isotopes ($^{39}$K and $^{41}$K), and a ground-state crossover resonance in saturated absorption.  Such resonances occur for atoms whose Doppler shifts equal half the frequency difference between transitions from different ground states.  In this case hyperfine pumping \cite{Smi04} causes increased absorption, corresponding to an inverted feature.  Since the ground state splittings of rubidium and cesium are an order of magnitude greater than for potassium, this velocity class is essentially unpopulated and ground-state crossovers are not observed.  An example potassium saturated absorption spectrum is shown in Fig.~\ref{fig:01}.  Due to its large natural abundance ($93.3\%$) the $^{39}$K features are dominant.  The peak on the red side of the Doppler background comprises all of the $F=2\rightarrow F^\prime=3$ transitions and associated excited-state crossovers, and the dip near the center is from the ground-state crossovers (here $F$ is the total electronic plus nuclear angular momentum).  We will refer to these simply as the $F=2$ and crossover (CO) features for brevity.  As the vapor pressure of potassium is lower than those of rubidium and cesium, it is beneficial to heat the cell.  Throughout this work the cell was maintained at $45$--$48~^\circ$C, corresponding to a maximum Doppler-broadened absorption of $35$--$40\%$. More detailed accounts of the D$_2$ level structure and the optimization of saturated absorption are given in \cite{Mud12}.

The rest of this paper is organized as follows.  In Section 2 we study polarization spectroscopy.  A comparison is made between the use of a beam splitter and a mirror for aligning the counter-propagating pump and probe beams.  We then study the dependence of the error signal slope and amplitude on the pump and probe intensities.  In Section 3 we discuss magnetic dichroism.  In our experiment, the dispersive features obtained in the conventional way were too small to be useful for laser locking, leading us to introduce two alternative variations.  Of these, one produces a steeper, larger, and more symmetric signal, and we optimize this method through its dependence on magnetic field and pump and probe intensities.  Finally, in Section 4 we discuss our results, and compare our modulation-free methods with our previous work.

\section{Polarization spectroscopy}
\begin{figure}[htbp]
\centering\includegraphics[width=8cm,clip=true]{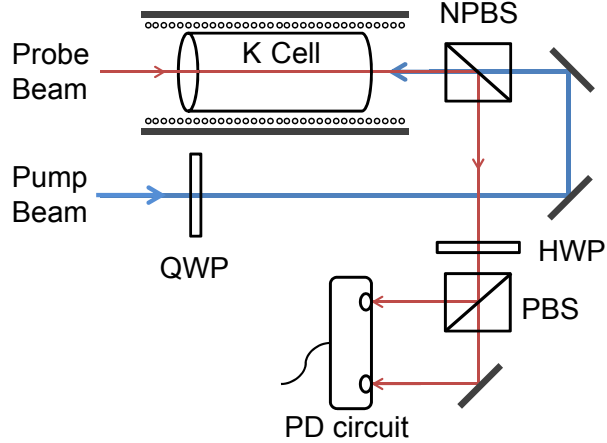}
\caption{\label{fig:02}Schematic setup for polarization spectroscopy.  Pump and probe beams are derived from a single incident beam (not shown).  The potassium reference cell is wrapped in a foil heater and a solenoid, and then placed in a magentic shield.  NPBS: non-polarizing (50:50) beam splitter; QWP: quarter-wave plate; HWP: half-wave plate; PBS: polarizing beam splitter; PD: photodiode.}
\end{figure}

Polarization spectroscopy (PS) employs a circularly polarized pump beam to induce an anisotropic distribution of the atomic Zeeman sub-states.  A simplified schematic of our setup is shown in Fig.~\ref{fig:02}.  A solenoid around the cell produces a weak ($\sim 20$~mG) uniform magnetic field along the optical axis, which sets the quantization axis for the atoms. If this field points in the direction of the pump beam propagation, then a left-hand circularly polarized pump will drive $\sigma^+$ transitions, which tend to transfer atoms to high-$m_F$ Zeeman sublevels \cite{Ber11}. Here $m_F$ is the quantum number associated with the projection of total (nuclear plus electronic) angular momentum.  The probe beam is linearly polarized, corresponding to a coherent superposition of left- and right-hand circular polarizations, which experience different phase shifts and absorptions due to the pump-induced birefringence and dichroism, respectively. In the original demonstration of polarization spectroscopy, a single detector was used, and the probe was analyzed with a pair of nearly crossed polarizers \cite{Wie76}.  This allows resonances to be observed on a very small background, but does not produce a zero-crossing error signal as required for laser locking. Pearman \textit{et al.} introduced the analyzer used here \cite{Pea02}, which consists of a half-wave retarder and polarizing beam splitter (PBS); the outputs of the PBS are detected and subtracted with a photodiode differencing circuit.  

If the incident probe beam polarization is set to $45~^\circ$ from the PBS axis, then in the absence of any anisotropy the difference signal is zero.  In the presence of the $\sigma^+$ pump beam, a re-population of the ground states favoring larger $m_F$ takes place.  The two circularly polarized components of the probe then experience different absorption coefficients and refractive indices, corresponding to circular dichroism and birefringence, respectively.  The dichroism, characterized by the difference of absorption coefficients $\Delta\alpha=\alpha_+-\alpha_-$, causes the probe to become elliptically polarized; the birefringence, characterized by the difference of refractive indices $\Delta n=n_+-n_-$ causes a rotation of the polarization axis.  For a weak probe intensity (\textit{i.e.}, below atomic saturation) and neglecting any birefringence from the cell windows, the detected difference signal for an isolated resonance is given by \cite{Pea02}
\begin{eqnarray}
\Delta V = V_0\,e^{-\alpha L}\left(\Delta\alpha_0\,L\,\frac{x}{1+x^2}\right)
\label{eq:PS}
\end{eqnarray}
where $x=(\omega-\omega_0)/\Gamma$ is the normalized detuning from the resonance $\omega_0$, and $\Gamma$ is the transition line width, which may be greater than $\gamma$ due to power and transit-time broadening or additional dephasing effects. The amount of probe absorption is dictated by the average absorption coefficient $\alpha=(\alpha_++\alpha_-)/2$ and the cell length $L$, which was 75~mm throughout this work.  The magnitude of the anisotropy is characterized by $\Delta\alpha_0$, the value of $\Delta\alpha$ on resonance. The voltage $V_0$ is a function of the incident probe power and the transimpedance of the photodiode circuit. The home-built circuit used in this work amplifies the current difference between two photodiodes with $100~{\rm k}\Omega$ transimpedance and $4\times$ additional voltage gain; the design 3~dB bandwidth is 750~kHz. From Eq.~(\ref{eq:PS}), we expect a dispersive signal which goes through zero at resonance and again at large detunings on either side.

\begin{figure}[htbp]
\centering\includegraphics[width=10cm]{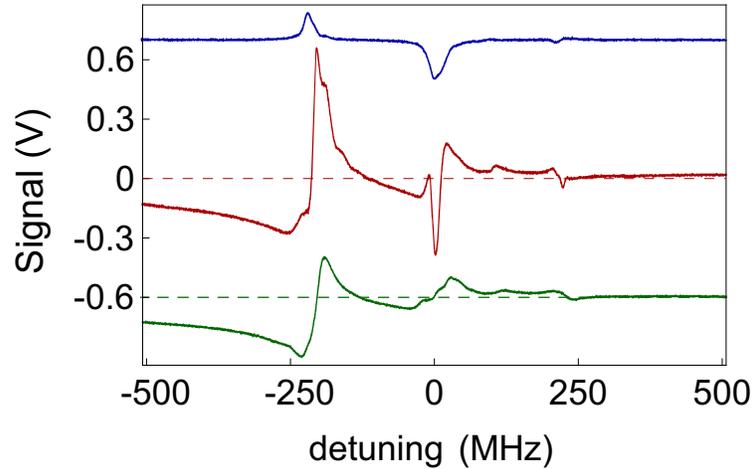}
\caption{\label{fig:03}Polarization spectroscopy of potassium.  The middle curve is a PS spectrum taken with the setup show in Fig.~\ref{fig:02}, and the lower curve is the spectrum obtained under the same conditions when the NPBS is replaced with a gold mirror, resulting in a $\sim 40$~mrad crossing angle between pump and probe beams. For comparison, the upper curve shows saturated absorption in a separate cell, with the Doppler background subtracted.  The curves have been shifted vertically for clarity.}
\end{figure}

An example polarization spectrum is shown in Fig.~\ref{fig:03}, with a saturated absorption spectrum for reference.  The PS pump intensity was $5.6\pm 1.0$~mW/cm$^2$ and the probe was $3.8\pm 0.7$~mW/cm$^2$, where the systematic uncertainty in intensity is due to the beam size calibration, as described in \cite{Mud12}. The error signal shows numerous overlapping transitions but is still suitable for laser locking.  The $F=2$ feature is much larger than the crossover, probably due to the dominance of the closed $F=2\rightarrow F^\prime=3$ transition \cite{Pea02}.  In building the setup shown in Fig.~\ref{fig:02}, we chose to combine pump and probe on a non-polarizing (50:50) beam splitter (NPBS).  This allows perfect overlap of the pump and probe beams, but deviations from ideal polarization-independence could distort the signal.  We were especially concerned about this since the beam splitters at hand were designed for the wavelength range 450--700~nm.  As a test we replaced the NPBS with a gold mirror, resulting in an estimated crossing angle between beams of 40~mrad.  The results are shown in Fig.~\ref{fig:03}, under the same experimental conditions as with the NPBS.  The spectrum obtained with the NPBS has larger amplitude and more sub-structure.  Specifically, the crossover effectively disappears when using the mirror.  A finite crossing angle presents two potential problems.  First there is a residual Doppler broadening.  A straight-forward, though tedious, calculation gives an estimated broadening $\Delta_{\rm eff}=(\theta/2)\Delta$ where the single-beam Doppler width $\Delta=2\pi\times 390$~MHz for $^{39}$K at 300~K.  For $\theta=0.04$, this gives 8~MHz, which is significant in light of the close-packing of resonances in potassium.  Additionally, the volume of gas in the overlap region between pump and probe is reduced.  For our $\theta$ and $L$ this effect is expected to be negligible.  Based on the results of Fig.~\ref{fig:03}, we used the NPBS for all remaining measurements.

\begin{figure}[htbp]
\centering
	\parbox{0.5\textwidth}{\includegraphics[width=6cm]{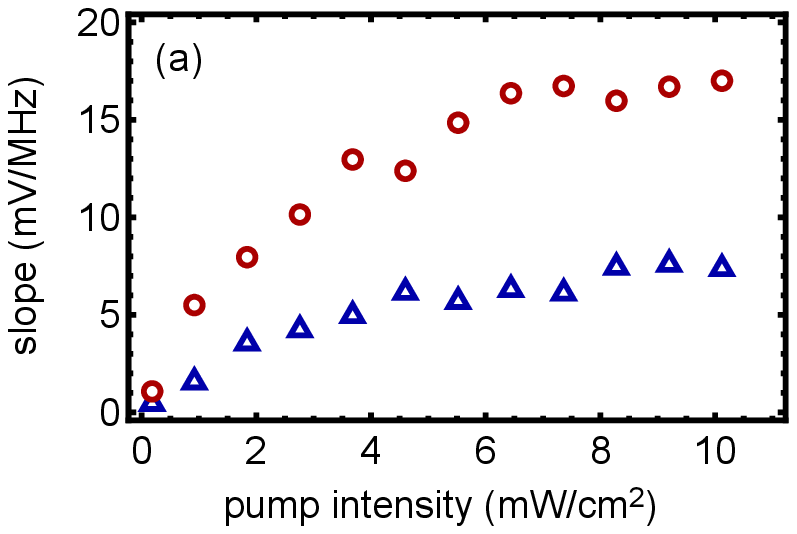}}\hfill
	\parbox{0.5\textwidth}{\includegraphics[width=6cm]{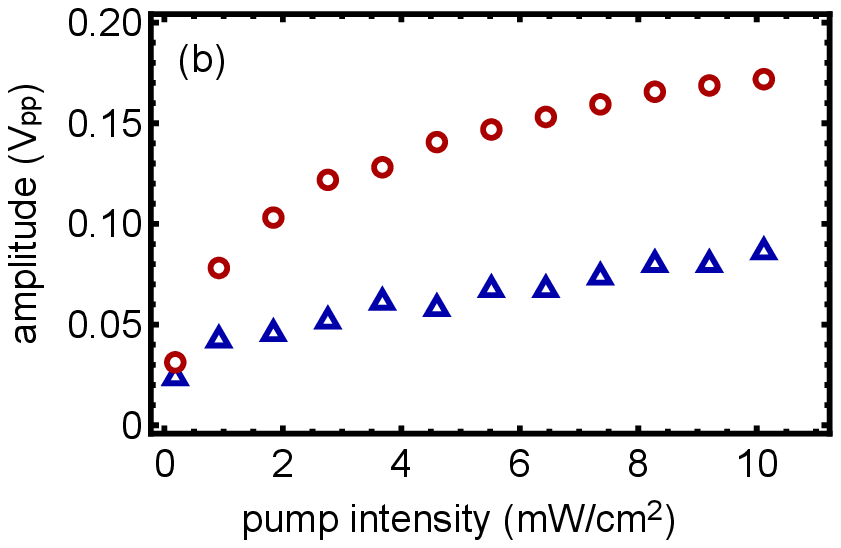}}
\caption{\label{fig:04}Dependence of polarization spectroscopy on pump intensity, with fixed $1.9\pm0.3$~mW/cm$^2$ probe intensity.  (a) Slope at the zero crossing.  Red circles correspond to the $F=2$ feature, and blue triangles to the crossover.  (b) Peak-to-peak amplitude between turning points. }
\end{figure}

We vary the polarization spectrum slope and peak-to-peak amplitude by individually varying the pump and probe intensities.  For small intensities, $V_0$ is expected to be linear with probe intensity, and $\Delta\alpha_0$ linear with pump. Assuming no power broadening, the signal amplitude and slope should then depend linearly on either intensity.  For higher intensities, power broadening should lead to an increase in $\Gamma$, which would cause the signal amplitude to saturate and the slope to roll over.  In a two-level approximation without transit-time effects, the relevant scale is set by the saturation intensity $I_{\rm sat}$ which is equal to $1.75$~mW/cm$^2$ for the $F=2\to F^\prime=3$ cycling transition \cite{Wan97,Fal06}.  In Fig.~\ref{fig:04} we vary the pump intensity with the probe fixed at $1.9\pm0.3$~mW/cm$^2$. The crossover typically has about half the amplitude and slope of the $F=2$ feature.  The latter reaches its half-maximum amplitude below 2~mW/cm$^2$, consistent with expectations if it is dominated by the cycling transition.  There is no clear saturation of the crossover amplitude.  However this feature does not arise from a simple two-level resonance.  In fact it comprises a number of nearly overlapping three- and four-level systems.  A full analysis of the saturating behavior of the crossover is outside the scope of this work.  Modifications to $I_{\rm sat}$ in real systems have been described in detail elsewhere (see, for example, \cite{Pap80,She09}).

\begin{figure}[htbp]
\centering
	\parbox{0.5\textwidth}{\includegraphics[width=6cm]{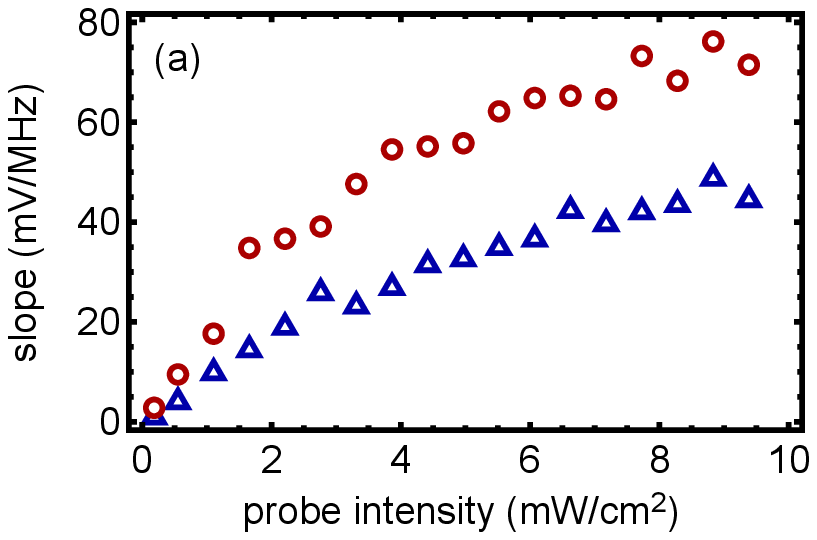}}\hfill
	\parbox{0.5\textwidth}{\includegraphics[width=6cm]{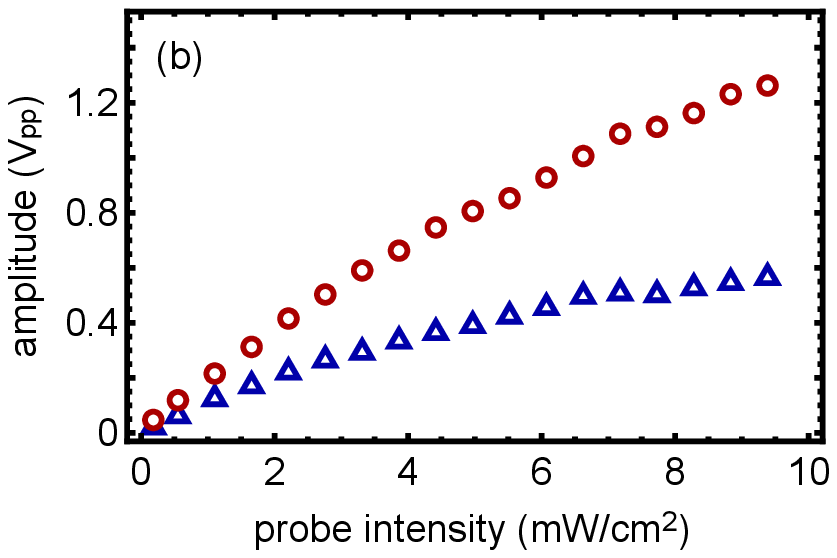}}
\caption{\label{fig:05}Dependence of polarization spectra on probe intensity, with fixed $5.6\pm1.0$~mW/cm$^2$ pump intensity. (a) Slope at the zero crossing.  Points are as in Fig.~\ref{fig:04}.  (b) Peak-to-peak amplitude.}
\end{figure}

The effects of varying probe intensities are shown in Fig.~\ref{fig:05}.  The amplitude data show only weak saturation with probe intensity.  Remarkably, this persists even when the probe intensity is greater than the pump and several times greater than the nominal (\textit{i.e.}, cycling) saturation intensity. However some power broadening does occur, as evidenced from the slope data, with the half-maximum slope reached at slightly higher intensities than above.  As our primary interest is to obtain an error signal with large amplitude and slope for laser locking, we simply note that polarization spectroscopy of potassium appears to be very robust to high intensities, especially around the $F=2$ feature.

\section{Magnetically-induced dichroism}
Magnetically-induced dichroism (MD) uses a longitudinal magnetic field to Zeeman-shift positive and negative $m_F$ states in opposite directions, causing a frequency-dependent circular dichroism in the medium.  A linearly polarized probe is decomposed into $\sigma^\pm$ components whose transmitted powers are subtracted.  If the magnetic field strength is chosen to produce a Zeeman shift on the order of the absorption line width, a dispersive shape is obtained.  In the context of laser locking, magnetic dichroism was introduced by Corwin \textit{et al.} using Doppler-broadened absorption and thus relatively large ($\sim 100$~G) fields \cite{Cor98}.  The resulting advantage is an extremely large capture range, which is on the order of the Doppler width (100s of MHz).  Unfortunately, in our experience with rubidium this comes at the price of stability, with frequency drifts of at least a few MHz.  If robustness to perturbations is less important than reduced fluctuations, one can introduce a linearly polarized pump beam, resulting in sub-Doppler magnetic dichroism.  This has the added advantage of requiring significantly smaller fields, since the Zeeman shift should now be on the order of the natural, rather than Doppler, line width.  Sub-doppler magnetic dichroism appears to have been first presented by Shim \textit{et al.} using the $^{85}$Rb D$_2$ transitions \cite{Shi99}.  It was shown that a small field-induced shift (from a 3~G field) results in a dispersive feature approximately proportional to the derivative of the saturated absorption signal.  This work seems to have been largely overlooked; the technique was independently re-discovered by two groups a few years later using the D$_1$ transitions of $^{23}$Na and $^{85}$Rb \cite{Was02} and the D$_2$ transitions of $^{85}$Rb and $^{87}$Rb.  By slightly increasing the magnetic field ($\sim 10$~G) it was possible to increase the error signal slope.

When we tried sub-Doppler magnetic dichroism with potassium, we were unable to obtain a suitable locking signal.  When the probe polarization was set to obtain zero volts away from resonance, the Doppler background overwhelmed the dispersion and no zero-crossing occurred.  If instead the polarization was set to obtain a flatter background, the offset was too large.  Since preparing this manuscript, we have become aware of one study of sub-Doppler magnetic dichroism in potassium \cite{Pic12}.  Although the Doppler background was too large to reliably lock near the $F=2$ line, the crossover was found to work well.  We are unsure why our experiment failed to obtain similar results, but we believe the difference may be in the wave plates we use, which are designed for 780~nm and were chosen for availability and cost.  Probe polarization impurities have been shown to cause large offsets in rubidium MD spectra \cite{Har08}.  Additionally, we heat our cell to increase the vapor pressure, which may cause increased birefringence of our cell windows.  No heating of the cell was reported in \cite{Pic12}.

\begin{figure}[htbp]
\centering
	\parbox{0.8\textwidth}{\includegraphics[width=10cm]{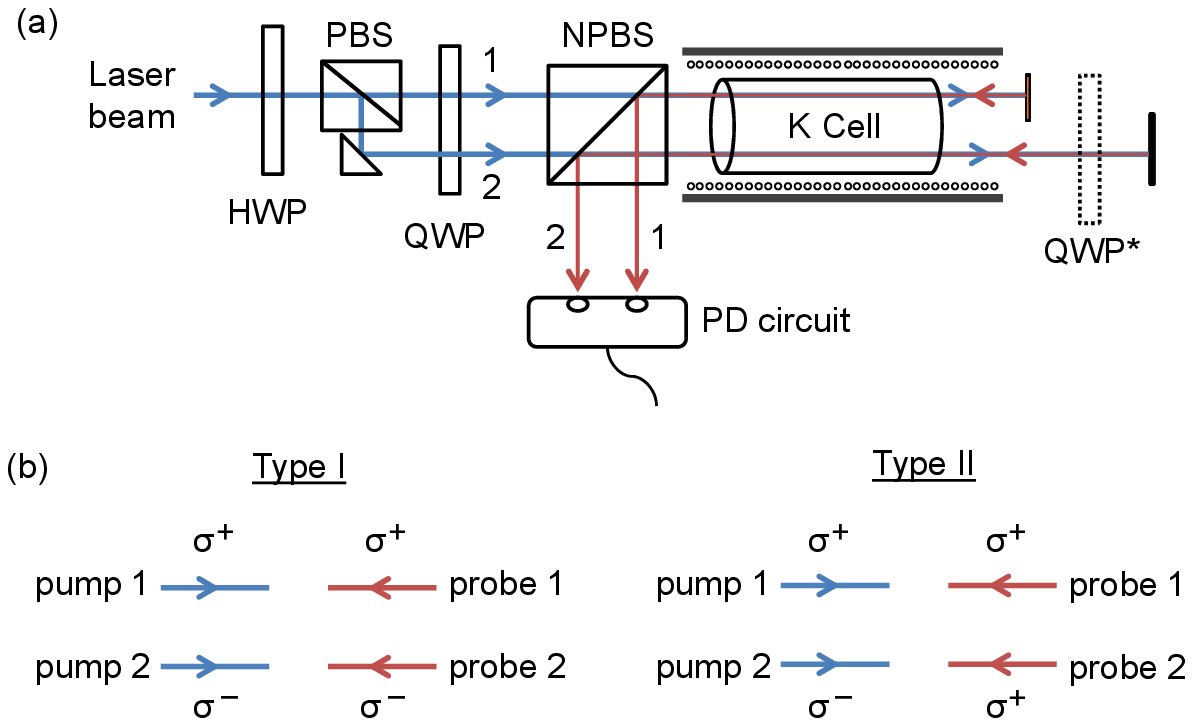}}
	\parbox{0.5\textwidth}{\hspace{1.3cm}\includegraphics[width=5cm]{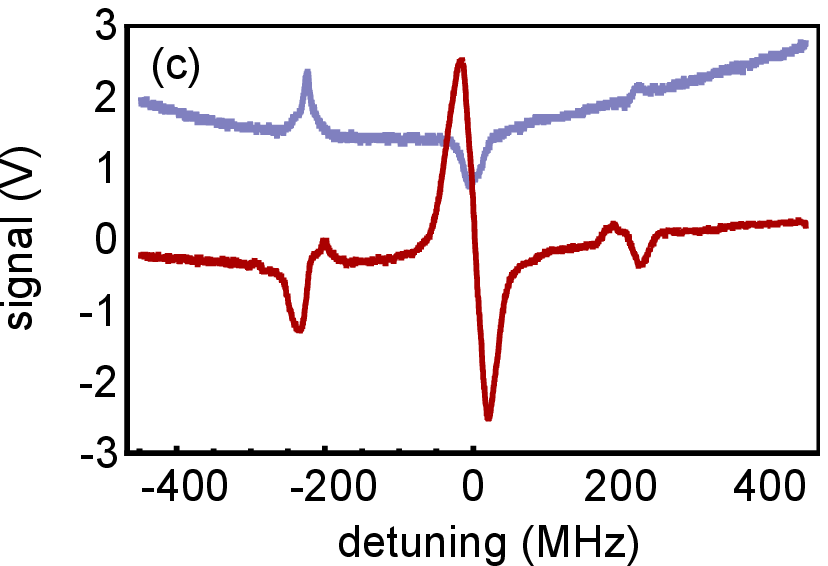}}\hfill
	\parbox{0.5\textwidth}{\includegraphics[width=5cm]{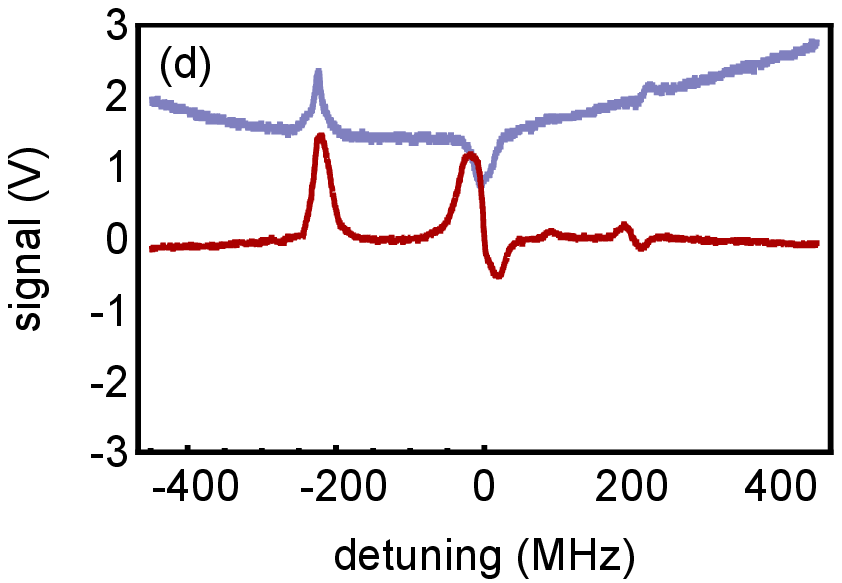}}
\caption{\label{fig:06}Split-beam magnetically induced dichroism. (a) Simplified schematic.  Notations are as in Fig.~\ref{fig:02}.  The dashed quarter-wave plate QWP* is absent in Type I spectroscopy and present in Type II. (b) Comparison of pump and probe polarizations for Type I and II configurations. (c) Type I spectroscopy.  The dark (red) curve shows the magnetically induced dichroism, and the faint (blue) curve shows saturated absorption for reference.  For the MD signal, the pump intensity per beam was $7.3\pm1.5$~mW/cm$^2$ (measured just before the cell) and the magnetic field was 10.3~G. (d) Type II spectroscopy, under the same conditions as in (c).}
\end{figure}

There are two straightforward ways to improve the magnetic dichroism signal.  One can either increase the atomic anisotropy, or reduce the Doppler background. To this end, we consider two spatially separated pairs of pump and probe beams, as shown schematically in Fig.~\ref{fig:06}(a).  In conventional single-beam dichroism, pump and probe are both linearly polarized.  By symmetry, this produces a roughly uniform population of $m_F$ states, which results in an average line strength less than the maximum (cycling) transition.  Here we increase the dichroism by using separated $\sigma^\pm$ pump beams, each of which transfers atoms into larger-$|m_f|$ states.  If each probe beam has the same polarization as the corresponding pump, one obtains stronger absorption than in the usual case.  We will refer to this configuration of pump-probe polarizations as Type I. On the other hand, the background is given by the differential Doppler absorption. This can be reduced ideally to zero by having probe beams which are both of the same polarization, say $\sigma^+$, at the expense of an asymmetric dispersion due to the different Clebsch-Gordan coefficients seen by the two probes.  This is summarized schematically in Fig.~\ref{fig:06}(b), and example spectra are shown in Fig.~\ref{fig:06}(c) and (d).

As can be seen in Fig.~\ref{fig:06}, type II spectroscopy produces only a slightly flatter background, with relatively small dispersive features.  In contrast, the type I configuration produces a larger, more symmetric error signal, especially at the crossover.  We therefore restrict the rest of discussion to the type I crossover.  The relative prominence of the crossover suggests that hyperfine pumping plays a significant role \cite{Smi04}.  To see this, consider atoms with zero velocity along the optical axis, for which the pump and probe are at the same frequency and therefore address the same hyperfine ground state.  Then hyperfine pumping results in a transfer of population into the other ground state, which is out of resonance with the probe.  In contrast, for ground state crossovers, this process increases the population interacting with the probe (recall the enhanced CO absorption in Fig.~\ref{fig:01}).

To understand the basic physics of the Type I spectroscopy in more detail, we consider a pair of simple two-level systems, each corresponding to one of the pump-probe pairs.  For a probe below the saturation intensity, the signal at each photodiode is proportional to $\exp(-\alpha)$, where $\alpha$ is the relevant absorption coefficient.  In our configuration, the two probes experience absorption coefficients \cite{Ber11}
\begin{eqnarray}
\alpha_\pm = \alpha_{D\pm}(1-\mathcal{L}_\pm)
\label{eq:alphapm}
\end{eqnarray}
where $\alpha_{D\pm}$ are the Doppler-broadened absorption coefficients, and $\mathcal{L}_\pm=A_\pm/[1+(x\pm\delta)^2]$ are sub-Doppler Lorentzians whose amplitudes $A_\pm$ and widths generally depend on the pump and probe intensities.  As above $x$ is the detuning normalized to the broadened width; $\delta$ is the Zeeman shift similarly normalized.  For small $\alpha_\pm$ the difference signal is
\begin{eqnarray}
\Delta V \approx \alpha_- - \alpha_+
\end{eqnarray}
The amplitude $A_+=A_-$ by symmetry.  The Zeeman shift is on the order of $\Gamma$ which is much smaller than the Doppler width $\Delta$.  In this case $\alpha_{D+}\approx\alpha_{D-}$, and we can write
\begin{eqnarray}
\Delta V = A \left[ \frac{1}{1+(x-\delta)^2} - \frac{1}{1+(x+\delta)^2} \right]
\label{eq:MD}
\end{eqnarray}
For small applied magnetic fields, the signal is thus approximately proportional to the derivative of a Lorentzian \cite{Shi99}.

From Eq.~(\ref{eq:MD}) the slope at the lock point is $4A\delta/(1+\delta^2)^2$, which grows linearly with small $\delta$ and has a maximum for $\delta_0=3^{-1/2}\approx0.6$.  It is possible to obtain a closed expression for the capture range, defined by the peak-to-peak width of the error signal, as well as the signal amplitude.  However the expressions are complicated and not very enlightening.  In the regime of interest ($\delta<1$), the predicted capture range is approximately constant for small fields and only grows by $25\%$ by $\delta_0$.  In contrast, the error signal amplitude grows linearly for small $\delta$ and then saturates when the $\pm$ peaks fully separate.

\begin{figure}[htbp]
\centering
	\parbox{0.5\textwidth}{\includegraphics[width=6cm]{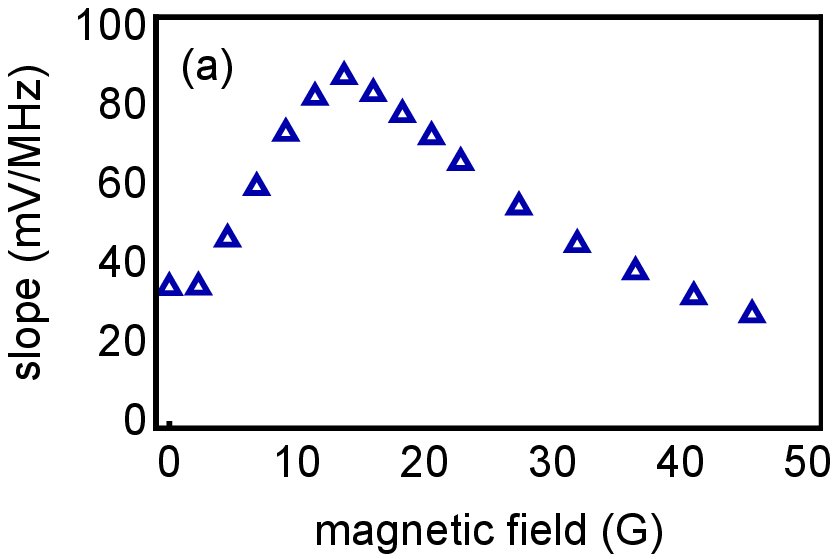}}\hfill
	\parbox{0.5\textwidth}{\includegraphics[width=6cm]{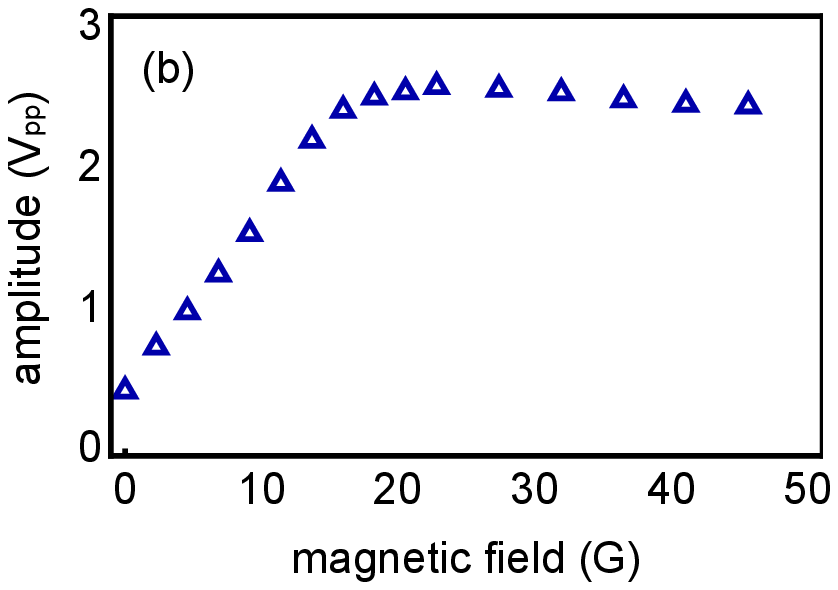}}
\caption{\label{fig:07}Dependence of the magnetic dichroism error signal with applied field. Only data for the crossover are shown.  The pump intensities were $2.7\pm 0.5$~mW/cm$^2$ each, measured just before the cell. (a) Slope at the zero-crossing. (b) Peak-to-peak amplitude.}
\end{figure}

Figure \ref{fig:07} shows the observed dependence of the crossover error signal on magnetic field.  Despite the complexity of the real atomic system, the behavior is qualitatively as expected from the two-level model.  We observe a small feature even at zero applied field, which we attribute to the residual ambient field and imperfect light polarization.  The optimum slope is obtained at $\sim 14$~G.  In the limit of perfect optical pumping into large-$|m_F|$ states, the crossover should be dominated by the $(F,|m_F|)=1\to (F^\prime,|m_F^\prime|)=2$ and $(F,|m_F|)=2\to (F^\prime,|m_F^\prime|)=3$ transitions, which have the greatest oscillator strengths \cite{Met99}.  In a magnetic field $B$, the resonances are Zeeman shifted in energy by $(g_F^\prime m_F^\prime-g_F m_F)\mu_BB$, where $\mu_B\approx9.28\times 10^{-28}$~J/G is the electron magnetic moment and $g_F$ is the hyperfine Land\'{e} $g$-factor. For the ground states $g_1=-1/2$ and $g_2=1/2$, and for the excited states $g_2^\prime=2/3=g_3^\prime$.  The optimum field thus corresponds to shifts of 19~MHz and 35~MHz for the $2-3^\prime$ and $1-2^\prime$ transitions, respectively.  Although this is much larger than the natural line width $\gamma/(2\pi)$, our saturated absorption crossovers are typically $\sim 20$~MHz wide \cite{Mud12}.

\begin{figure}[htbp]
\centering
	\parbox{0.5\textwidth}{\includegraphics[width=6cm]{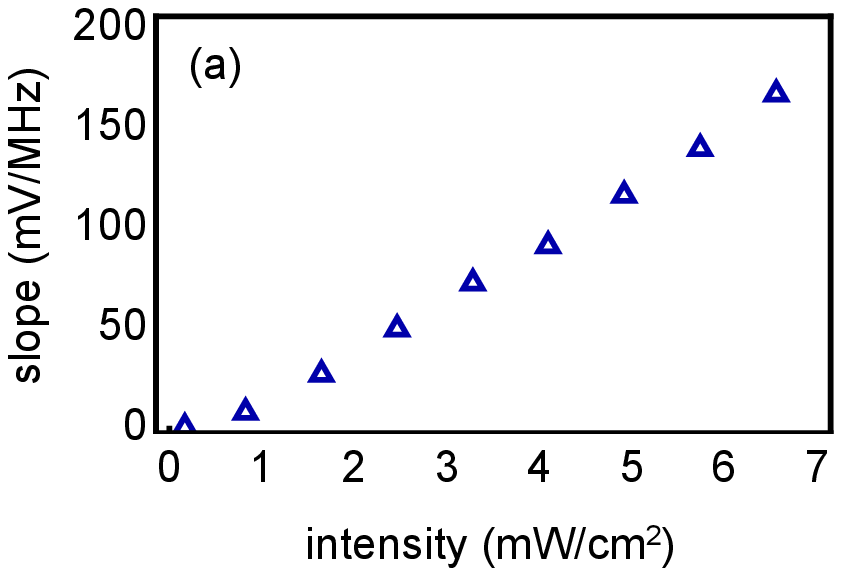}}\hfill
	\parbox{0.5\textwidth}{\includegraphics[width=6cm]{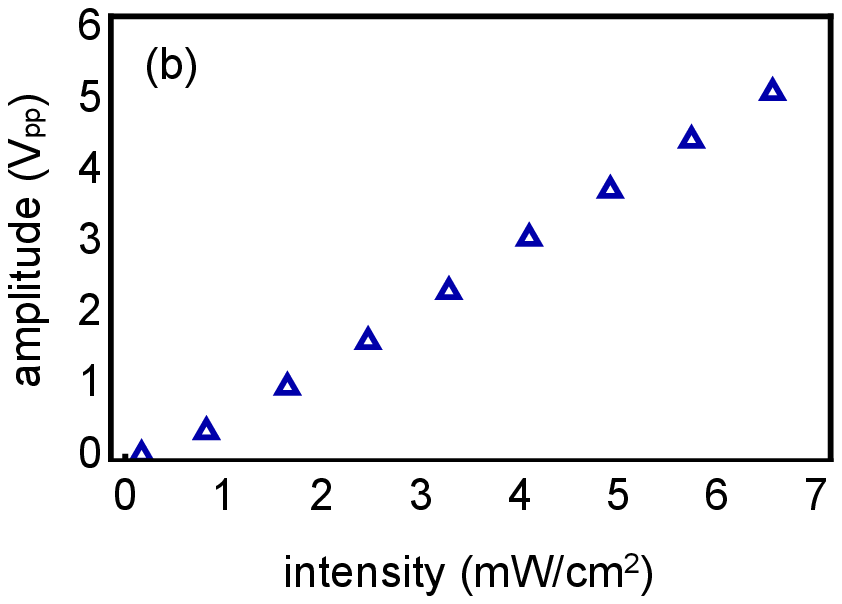}}
\caption{\label{fig:08}Dependence of the MD error signal on beam intensity.  Shown is the intensity per pump beam, measured just before the cell.  The magnetic field was $10.3$~G. (a) Slope at the zero-crossing. (b) Peak-to-peak amplitude.}
\end{figure}

In Fig.~\ref{fig:08} we show the dependence of the error signal on beam intensity. Note that since we retroreflect our pump beams, both pump and probe beam intensities are varying together. Above $I_{\rm sat}$ the amplitude and slope are essentially linear over the whole range of intensities we have tried.  The lack of saturation and the constant ratio of amplitude to width suggests there is little or no broadening, even at intensities well above $I_{\rm sat}$.

\section{Discussion}
We have studied polarization spectroscopy (PS) and magnetic dichroism (MD) for laser frequency stabilization near the potassium D$_2$ transitions at 767~nm.  These methods produce dispersive error signals which are roughly linear in detuning around a zero-voltage locking point. The main practical difference between the two methods is that the PS signal is best suited for locking near the $F=2$ feature, while the MD signal favors the crossover.  This reflects the general strength of PS lines at cycling transitions, whereas MD spectroscopy can be used for nearly any strong absorptive feature; this distinction was noted in the case of rubidium in \cite{Har08}.  The preferred lock point depends on the details of the experiment, such as which isotope is under study, the desired detuning from resonance, and whether or not frequency shifting elements such as acousto-optic modulators are to be used.  

As described in \cite{Mud12}, a critical parameter for laser locking is the error signal slope.  This represents a gain term within the total stabilization loop which is not constrained by limitations such as the fixed gain-bandwidth product typical of electronic components. In this sense, our modified MD signal appears to be superior to PS, giving a $\sim 2\times$ greater slope.  However, after preparing this manuscript, we became aware of a type of polarization spectroscopy using a split-beam configuration similar to that used in our MD experiments \cite{Tiw06}.  Such a modification might also increase the slope of our PS signal.  Other figures of merit include the signal amplitude and the capture range, which we take to be the peak-to-peak width.  The PS capture range is slightly larger, but the MD signal amplitude is $4\times$ greater.  In turn, the PS noise level was typically 7~mV (root-mean-squared), compared to 20~mV for MD.  Both methods tolerate relatively high intensities, with MD especially insensitive.  Although similar work with rubidium MD spectroscopy showed an optimum intensity \cite{Har08}, we were not able to reach high enough laser powers to see a roll-over in our setup.

Although our system has not been in operation long enough to acquire information on long-term stability, our previous experience with rubidium suggests that both methods will be sensitive to temperature-dependent drifts in polarization.  Furthermore, this experience suggests that the lock point does not occur exactly at resonance, but can be as much as 10~MHz off-resonance.  We do not yet know if this offset is stable over time, but we expect it to similarly drift with polarization and vapor pressure (through the cell temperature).  One possible advantage of magnetic dichroism may be with respect to $B$-field sensitivity.  Polarization spectroscopy is typically performed at zero or low bias field (we used 20~mG), whereas MD was optimized with 14~G.  We therefore expect MD spectroscopy to be more robust to changes in ambient field.

\begin{table*}
	\centering
	\caption{Comparison of modulation-based and modulation-free error signals.  Results for direct modulation (DM) and modulation transfer (MT) are from \cite{Mud12}.  Numbers for MD correspond to the crossover, and all others to the $F=2$ feature.}
		\begin{tabular}{cccccc}\hline
			Method & Slope & Amplitude & Capture Range & Noise & Bandwidth \\ \hline
			PS & 80~mV/MHz & 1.2~V$_{\rm pp}$ & 56~MHz & 7~mV$_{\rm rms}$ & 750~kHz \\
			MD & 170 & 5.1 & 36 & 20 & 750 \\ 
			DM & 100 & 0.4 & 15 & 15 & 200 \\ 
			MT & 600 & 1.3 & 5 & 15 & 200 \\ \hline			
		\end{tabular}
	\label{tab:compare}
\end{table*}

Finally, we compare our PS and MD results with our previous results using heterodyne spectroscopy.  In \cite{Mud12} we performed modulation spectroscopy either by direct phase modulation of the probe (DM) or by non-linear modulation transfer from the pump (MT).  The results are summarized in Table~\ref{tab:compare}.  The modulation-free methods presented here compare well with direct modulation, and use only standard optics available in most labs.  Modulation transfer is unique in both its high slope (600~mV/MHz) and narrow capture range (5~MHz).  This could be an advantage for experiments requiring a tight lock with relatively little error or drift in the lock-point.  However the exceedingly narrow capture range could cause the laser to lose its lock in noisy environments.

\section*{Acknowledgments}
This work was funded by EPSRC (EP/E036473/1).  We acknowledge useful discussions with members of the Cold Atoms group at the University of Birmingham.  We are grateful to V.~Boyer for pointing out the role of four-level ground-state crossovers, and to P.~Petrov and Y.~Singh for carefully reading the manuscript.

\end{document}